\documentstyle[psfig]{aipproc2}
\def\be{\begin{equation}}
\def\ee{\end{equation}}
\def\bea{\begin{eqnarray}}
\def\eea{\end{eqnarray}}

\begin{document}
\title{Removal of interference from external coherent signals}

\author{Alicia M. Sintes and Bernard F. Schutz}
\address{Max-Planck-Institut f\"ur Gravitationsphysik 
(Albert-Einstein-Institut),
Schlaatzweg 1, D-14473 Potsdam, Germany}

%National Center for Research\thanks{The National
%Center for Research is sponsored by the National
%Science Foundation.}, Boulder, Colorado 80307\\
%$^{\dagger}$National Standards Institute, Boulder, Colorado 11543}
%\lefthead{LEFT head}
%\rig436thead{RIGHT head}

\maketitle

\begin{abstract}
We present  a technique that we call coherent line removal, 
for  removing external coherent 
interference from  gravitational wave interferometer data.
We illustrate the usefulness of this technique applying it to the 
the data produced by the Glasgow laser interferometer in 1996 and
removing all those lines corresponding to the electricity supply frequency 
and its harmonics. We also find that this method seems to reduce the level
of non-Gaussian noise present in the interferometer and therefore, it
can raise the sensitivity and duty cycle of the detectors.

\end{abstract}

\section*{Introduction}

In the measured noise spectrum of the different  gravitational wave 
interferometer prototypes \cite{Abra,ga,J}, one observes peaks due 
external interference,
where the amplitudes are not stochastic in contrast to the stochastic
noise. The most numerous are powerline frequency harmonics.
In this paper we review how to  remove these very effectively
using a technique we call coherent line removal ({\sc clr}) \cite{AS1,AS2}.

 {\sc clr} is an algorithm able to remove interference present in the
 data while preserving the stochastic detector noise.  {\sc clr}
 works when the interference is present in many harmonics, as long as
   they remain 
 coherent with one another. Unlike other existing methods for removing
 single interference lines \cite{Th,p3}, {\sc clr} can remove the external
 interference without removing any \lq single line' signal buried by the
 harmonics. The algorithm works even when the interference frequency changes.
{\sc clr} can be used to remove all harmonics of periodic or 
broad-band signals (e.g., those which change frequency in time), even when 
there is no external reference source.  {\sc clr} requires little or
no a priori knowledge of the signals we want to remove. This is a 
characteristic that distinguishes it from other methods 
such as adaptive noise
cancelling \cite{anc}.
It is \lq safe' to apply this technique to gravitational wave data
because we expect that coherent gravitational wave signals will appear
with at most the fundamental and one harmonic \cite{Sc2}.
Lines with multiple harmonics must be of terrestrial origin.
 
In this paper, we illustrate the usefulness of this new technique by applying
it to the data produced by the Glasgow laser interferometer in March 1996
and removing all those lines corresponding to the electricity supply frequency 
and its harmonics. As a result the interference is attenuated or eliminated
by cancellation in the time domain and the power spectrum appears 
completely clean allowing the detection of signals that were 
buried in the
interference. Therefore, this new method appears to be good news as far
as searching for continuous waves (as those ones produced by pulsars
\cite{Sc2,T})
 is concerned.
 The removal improves the data in the time-domain as well. 
Strong interference produces a significant non-Gaussian component to the noise.
Removing it therefore improves the sensitivity of the detector to short
bursts of gravitational waves \cite{n}.

%\begin{figure}[t!] % fig 1
%\centerline{\epsfig{file=got.ps}}
%%\centerline{\epsfig{file=got.ps,height=3.5in,width=3.5in}}
%\caption{Figure captions are automatically centered and can be 
%more than one line.}
%\vspace*{10pt}
%\label{fig1}
%\end{figure}
%
%\begin{figure} 
%\centerline{\epsfysize=3.0in \epsfbox{figu1.ps} }
%\caption{Comparison of the structure of the lines at 250 Hz and at 350 Hz
%of the power
%spectrum of the Glasgow data.  
%The broad shape is due to the wandering of the incoming electricity 
%frequency.} 
%\label{phidot}
%\end{figure}

\section*{Coherent  Line Removal}

In this section, we summarize the principle of {\sc clr}.
For further details we refer the reader to \cite{AS2}.

We assume that the interference has the form
\be
y(t)=\sum_n a_n m(t)^n + \left( a_n m(t)^n\right)^* \ ,
\label{e3}
\ee
where $a_n$ are
complex amplitudes and  $m(t)$ is a nearly monochromatic function
 near
a frequency $f_0$.
The idea is to
 use the information in the different harmonics of the interference 
to construct a function $M(t)$ that is  as close a replica as possible of
$m(t)$ and then construct a function close to $y(t)$ which 
is subtracted from the output  of the system cancelling the interference.
The key is that real gravitational wave signals will not be present
with multiple harmonics and that $M(t)$ is constructed from many
frequency bands with independent noise. Hence, {\sc clr} will little
affect the statistics of the noise in any one band and
any gravitational wave signal masked by the interference can
be recovered without any disturbance.

We assume that the data produced by the system
is just the sum of the interference plus  noise
\be
x(t)=y(t)+n(t) \ ,
\label{e6}
\ee
where $y(t)$ is given by Eq.~(\ref{e3}) and the noise $n(t)$ in the
detector  is a zero-mean stationary
stochastic process.
The procedure consists in   defining a set of functions $\tilde z_k(\nu)$
in the frequency domain as
\be
\tilde z_k(\nu)\equiv \left\{
\begin{array}{cc}
\tilde x(\nu) & \nu_{ik}<\nu <\nu_{fk}\\
0 & \mbox{elsewhere}\ ,
\end{array}
\right.
\label{e8}
\ee
where  $(\nu_{ik}, \nu_{fk})$ correspond to the upper and lower frequency 
limits of the harmonics of the interference 
and $k$ denotes the harmonic considered.
 These functions are equivalent to
\be
\tilde z_k(\nu)= a_k \widetilde{m^k}(\nu) +\tilde n_k(\nu) \ ,
\ee
where $ \tilde n_k(\nu)$ is the noise in the frequency band of the
harmonic considered.  Their inverse  Fourier transforms yield
\be
z_k(t)=a_k m(t)^k +n_k(t) \ .
\ee
Since  $m(t)$ is supposed to be a narrow-band function near a frequency $f_0$,
each $z_k(t)$ is a  narrow-band function near $kf_0$. 
Then, we  define
\be
B_k(t)\equiv \left[ z_k(t)\right]^{1/k}\ ,\label{e10a}
\ee 
that can be rewritten as 
\be
B_k(t)= (a_k)^{1/k}m(t) \beta_k(t) \ , \qquad
\beta_k(t)=\left[ 1+ {n_k(t) \over a_k m(t)^k}\right]^{1/k} \ .
\label{e10}
\ee
All these  functions, $\{B_k(t)\}$, are almost monochromatic around the 
fundamental frequency, $f_0$, but they differ basically by a certain
complex amplitude. These factors, $\Gamma_k$, can easily be  calculated,
and  we can construct a set of functions   $\{b_k(t)\}$
\be
 b_k(t)=\Gamma_k B_k(t)\ ,
 \ee
such that, they all have the same mean value. Then,   $M(t)$ can be
 constructed  as a function of all $\{b_k(t)\}$
 in such a way
that it has the same mean and minimum variance. 
If 
$M(t)$ is linear with $\{b_k(t)\}$, the statistically the best is
\be
 M(t)=\left(\sum_k {b_k(t) \over {\rm Var}[\beta_k(t)]} \right) {\Big
 { /}}
\left( \sum_k {1 \over {\rm Var}[\beta_k(t)]}\right) \ ,
\ee
where
\be
{\rm Var}[\beta_k(t)]= {\langle n_k(t) n_k(t)^*\rangle\over  k^2
\vert a_k m(t)^k\vert^2}+ \mbox{corrections} \ .
\ee
In practice, 
we  approximate
\be
\vert a_k m(t)^k\vert^2 \approx \vert z_k(t)\vert^2 \ ,
\ee
and we assume  stationary noise. Therefore, 
\be
\langle n_k(t) n_k(t)^*\rangle= \int_{\nu_{ik}}^{\nu_{fk}} S(\nu) d\nu \ ,
\ee
where $S(\nu)$ is the power spectral density of the noise.

Finally, it only remains to determine the amplitude of the different 
harmonics, which can be obtained applying a least square method.

\section*{Removal of 50 Hz harmonics}
In this section, we present experimental results that demonstrate the
performance of the {\sc clr} algorithm and show its potential value. 
We apply this method to the data produced by the  Glasgow laser 
interferometer in March 1996 and the electrical interference is
successfully removed. 

In the study of the Glasgow data, we observe in the power spectrum many
 lines. Some of them are due to thermal noise (which we
will not consider here) and many others 
at multiples of 50 Hz due external interference, where the amplitudes
are not stochastic.
 In long-term Fourier transforms, the lines at multiples
of 50 Hz are broad, and the structure of different lines is similar 
apart from an overall scaling proportional to the frequency. In smaller
length Fourier transforms, the lines are narrow, with central frequencies
that change with time, again in proportion to one another. It thus appears
that all these lines are harmonics of a single source (e.g., the
electricity supply) and that their broad shape is due to the wandering of
the incoming electricity frequency.

\begin{figure}[b!]
\centerline{\vbox{ 
\psfig{figure=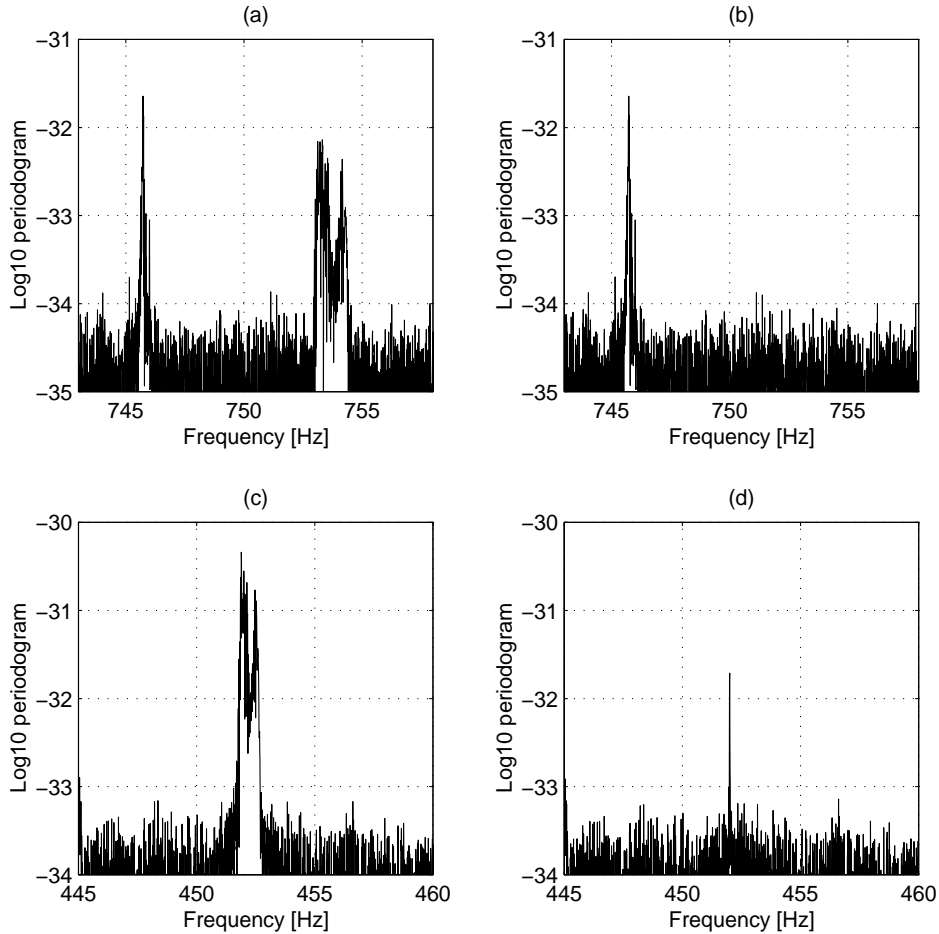,height=12.5cm,width=12.5cm} 
}} 
\vspace*{10pt}
\caption[]{Decimal logarithm of the
periodogram of $2^{19}$ points (approximately 2 minutes) of the Glasgow data. 
(a) One of the harmonics near 754 Hz.
(b) The same data after the removal of the interference as described 
in the text.
(c)  The same experimental  data
with an artificial  signal added at 452 Hz. 
(d) The data in (c) after the removal of the interference, revealing that
the signal remains detectable. Its amplitude is hardly changed by removing
the interference. 
} 
\vspace*{10pt}
\end{figure} 

In the Glasgow data, those lines at 1 kHz have a width of 5 Hz. Therefore,
we can ignore these sections of the power spectrum or we can try to 
remove this interference in order to be able to detect gravitational
waves signals masked by them.

In order to remove the electrical interference, we separate the data
into groups  of $2^{19}$ points (approximately two minutes) and, for each of 
them, the coherent line removal algorithm is applied. 
A detailed description  can be found in \cite{AS2}.

In Fig. 1, we show the performance of {\sc clr} on two minutes of data.
We can see how {\sc clr} leaves the spectrum completely clean of the
electrical interference  and keeps the intrinsic detector noise.
 {\sc clr} is also  applied to the true experimental data
with an external simulated signal at 452 Hz, that is initially hidden
due to its weakness and we succeed  in removing the electrical 
interference  while keeping the signal present in the data, obtaining
a clear outstanding peak over the noise level.

\begin{figure}[h!] 
\centerline{\vbox{ 
\psfig{figure=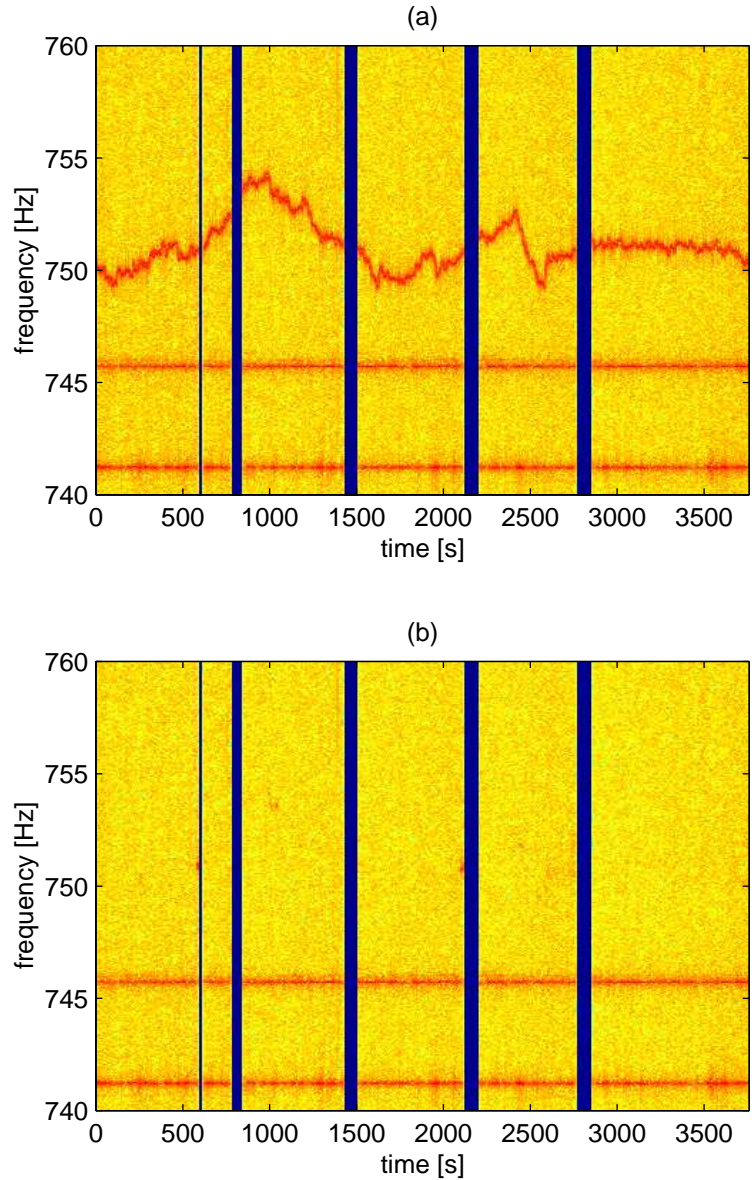,width=10.0cm}
}} 
\vspace*{10pt}
\caption[]{ Comparison of a zoom of the spectrogram.
The dark areas correspond to the periods in which the detector is
out of lock. (a) is obtained from the Glasgow data. We can observe
the wandering of the incoming electrical 
signal. The other two  remaining lines
at constant frequency correspond to violin modes. (b) The same spectrogram
as in (a) after applying coherent line removal, showing how the 
electrical interference is 
completely removed.
} 
\vspace*{10pt}
\end{figure} 
In Fig. 2, we compare a zoom of the spectrogram for the frequency
range between 740 and 760 Hz. There we can see the performance of the
algorithm on the whole data stream. We show
 how a line due to an harmonic of the electrical
interference in the initial data is removed.

We are interested in studying possible side effects of the line removal
on the statistics of the noise in the time domain. We observe that
the mean value is hardly changed. By contrast, a big difference is obtained
for the standard deviation. For the Glasgow data, its value is around
1.50 Volts. After the line removal, the standard deviation is reduced,
obtaining a value around 1.05 Volts. This indicates that a huge amount
of power has been removed. 

Further analysis reveals that values of
skewness and kurtosis are getting closer to zero after the line removal.
Values of skewness and kurtosis 
near  zero suggest a Gaussian nature. Therefore, we are
interested in studying the possible reduction of the level of non-Gaussian 
noise. To this end, we take a piece of data and we study
their histogram, calculating the number of events that lie between
different equal intervals. If we plot the logarithm of the 
number of events versus  $(x-\mu)^2$, where $x$ is the central position of
the interval and $\mu$ is the mean, in case of a Gaussian distribution,
all points should fit on a straight line of slope $-1/2\sigma^2$, where
$\sigma$ is the standard deviation. We observe that this is not the
case (see figure 3). Although, both distributions  seem to have
a linear regime, they present a break and then a very heavy tail.
The two distributions are very different. This is mainly due to the
change of the standard deviation.

\begin{figure}[h!]
\centerline{\vbox{ 
\psfig{figure=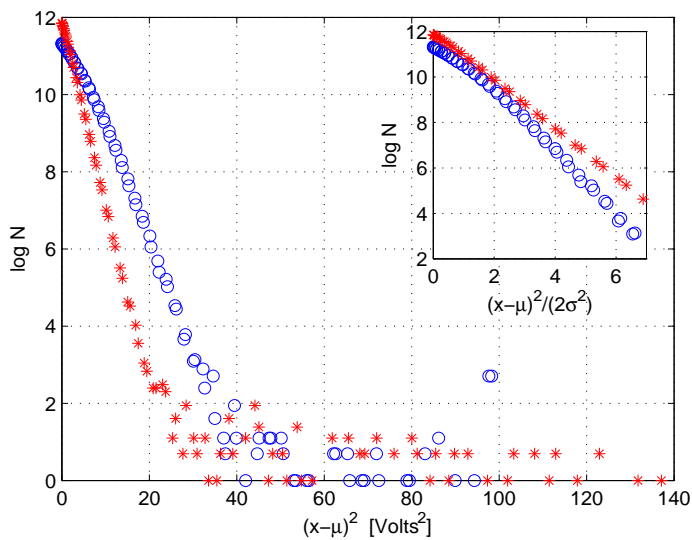,width=10cm} 
}} 
\vspace*{10pt}
\caption[]{ Comparison of the logarithm  plot of the histogram 
for $3.2\times 2^{19}$ points 
as a function of 
$(x-\mu)^2$. The circles correspond to the Glasgow data
and the stars to the same data after removing the electrical interference.
 The Glasgow data is 
characterized by $\mu=-0.0182$ Volts and $\sigma = 1.5151$ Volts.
After the line removal, we obtain the 
values of  $\mu=-0.0182$ Volts and $\sigma = 1.0449$ Volts.
In the right-hand corner, there is zoom  of the original figure, but
rescaled so that
the abscissa corresponds to $(x-\mu)^2/2\sigma^2$.
If the data  resembles a Gaussian distribution, 
we will expect a single straight line of slope -1. This is not the
case for the Glasgow data, but it seems to be satisfied  for the
clean data up to  $4\sigma$. The large number of points in the 
highest bin of the Glasgow data is an effect of saturating the ADC. These
points are spread to higher and lower voltages by line removal.
} 
\vspace*{10pt}
\end{figure} 

We can zoom the \lq linear' regime and change the scale in 
the abscissa  to $(x-\mu)^2/(2\sigma^2)$. Then, any Gaussian
distribution should fit into a straight line of slope -1. 
We observe that after removing the interference, it follows a 
Gaussian distribution quite well up to $4\sigma$. 
The original Glasgow data does not fit   a straight line anywhere.

In order  study the Gaussian character, we have also applied
 two statistical tests to the
data: the chi-square test  that measures  the discrepancies
between binned distributions,  and the one-dimensional Kolmogorov-Smirnov
test  that measures the differences between  cumulative 
distributions of a continuous data.

We computed the significance  probability for every $2^{12}$ points
of the data
using both tests and we checked whether the distribution are Gaussian or not.
The two tests are not equivalent but in any case, the values of the
significance probability would be close to unity for 
distributions resembling a Gaussian distribution. 
In both tests,  the significance  probability increased
after removing the electrical interference, showing  that this
procedure suppresses some non-Gaussian noise, although, generally 
speaking, the distribution was still 
non-Gaussian in character, presumably because of the heavy
tails, which are not affected by line removal. See \cite{AS2} for details.

\section*{Acknowledgments}
We would like to thank C. Cutler, A. Kr\'olak, M.A. Papa and
A. Vecchio for helpful discussions,
and  J. Hough and the gravitational waves group at Glasgow University
for providing their gravitational wave interferometer data for analysis.
This work was partially supported by the European Union, 
 TMR Contract
No. ERBFMBICT972771.


\begin{references}

\bibitem{Abra} Abramovici A., Althouse W., Camp J., Durance D.,
Giaime J.A., Gillespie A., Kawamura S., Kuhnert A.,
Lyons T., Raab F.J., Savage Jr. R.L., Shoemaker D., Sievers L.,
Spero R., Vogt R., Weiss R., Whitcomb S., Zucker M.,
{\it Physics Letters A }  {\bf 218}, 157  (1996).

\bibitem{ga} Maischberger K., Ruediger A., Schilling R., Schnupp L.,
Winkler W., Leuchs G.,  \lq\lq Status of the Garching 30 meter
prototype for a large gravitational wave detector", eds. 
 Michelson P.F., En-Ke Hu \&  Pizzella G., in {\it Experimental
Gravitational Physics}. World Scientific, Singapore, 1991, pp. 316-21.

\bibitem{J} Jones G.S., 1996, {\em Fourier analysis of the data produced
by the Glasgow laser interferometer in March 1996}, internal report,
Cardiff University, Dept. of Physics and Astronomy.


\bibitem{AS1} Sintes A.M.,  Schutz B.F.,  \lq\lq Removal of 
interference from gravitational wave spectrum",
in {\it proceedings of the 2nd workshop on Gravitational Wave Data Analysis}.
Orsay, France, 1998.

\bibitem{AS2} Sintes A.M.,  Schutz B.F., 
\lq\lq Coherent Line Removal: A new technique to remove interference from
the gravitational wave spectrum", 1998, to be published.


\bibitem{Th} Thomson D.J., 1982, \lq\lq Spectrum Estimation and 
Harmonic Analysis", in {\it Proceedings of the IEEE}, {\bf 70}, 1055-96 (1982).

\bibitem{p3} Percival D.B., Walden A.T., {\it Spectral
analysis for physical applications}, first edition, Cambridge 
University Press, 1993.

\bibitem{anc} Widrow B., Glover J.R., McCool J.M., Kaunitz J.,
Williamns C.S., Hearn R.H., Zeidler J.R., Dong E., Goodlin R.C., 
\lq\lq Adaptive Noise Cancelling: Principles and Applications", in
 {\it Proceedings of the IEEE}, {\bf 63}, 1692-1716 (1975).
 

\bibitem{Sc2} Schutz B.F.,  \lq\lq Detection of Gravitational Waves",
 in
{\em Relativistic Gravitation and Gravitational Radiation},
Marck J.A., Lasota J.P., eds., Cambridge University Press, 1997.


\bibitem{T} Thorne K.S.,  eds. Kolb E.W. \& Peccei R., in
{\it Proceedings of Snowmass 1994 Summer Study on Particle and Nuclear 
Astrophysics and Cosmology}. World Scientific, Singapore, 1995, p. 398.


\bibitem{n} Nicholson D., Dickson C.A., Watkins W.J.,
Schutz B.F., Shuttleworth J., Jones G.S.,
Robertson D.I., Mackenzie N.L., Strain K.A., Meers B.J., Newton G.P.,
Ward H., Cantley C.A., Robertson N.A., Hough J.,
Danzmann K., Niebauer T.M., R\"udiger A., Schilling R., Schnupp L., 
Winkler W., 
{\it  Physics Letters A} {\bf 218}, 175-180 (1996).


\end{references}
\end{document}